\documentstyle[epsfig]{elsart}
\newcommand{\lsim}{\mathrel{\rlap{\lower4pt\hbox{\hskip0pt$\sim$}}
\raise1pt\hbox{$<$}}}
\newcommand{\gsim}{\mathrel{\rlap{\lower4pt\hbox{\hskip0pt$\sim$}}
\raise1pt\hbox{$>$}}}
\newcommand{\sfrac}[2]{\mbox{\footnotesize $\frac{#1}{#2}$}}
\begin{document}
\begin{frontmatter}
%_______________________ Title, Authors ____________________________________
\hspace*{\fill}{Preprint Numbers: \parbox[t]{44mm}{ANL-PHY-8881-TH-98\\
                                MPG-VT-UR 121/98}}

\title{A dynamical, confining model and hot quark stars}
\author[ur]{D.~Blaschke,}
\author[uy]{H.~Grigorian,}
\author[uy]{G.~Poghosyan,}
\author[anl]{C. D. Roberts}
\author[ur,anl]{and S.~Schmidt}
\address[ur]{Fachbereich Physik, Universit\"at Rostock,
Universit\"atsplatz 1,\\ D--18051 Rostock, Germany}
\address[uy]{Department of Physics, Yerevan State University, Alex
Manoogian Str. 1,\\ 375025 Yerevan, Armenia}
\address[anl]{Physics Division, Bldg. 203, Argonne National
     Laboratory,\\ Argonne IL 60439-4843, USA}
\begin{abstract}
        We explore the consequences of an equation of state (EOS) obtained in
  a confining Dyson-Schwinger equation model of QCD for the structure and
  stability of nonstrange quark stars at finite-$T$, and compare the results
  with those obtained using a bag-model EOS.  Both models support a
  temperature profile that varies over the star's volume and the consequences
  of this are model independent.  However, in our model the analogue of the
  bag pressure is $(T,\mu)$-dependent, which is not the case in the bag
  model.  This is a significant qualitative difference and comparing the
  results effects a primary goal of elucidating the sensitivity of quark star
  properties to the form of the EOS.
\end{abstract}
\begin{keyword}
Dyson-Schwinger equations; Confinement; Dynamical chiral symmetry breaking;
Bag model; Quark matter equation of state; Quark stars\\[2mm]
\end{keyword}
\end{frontmatter}
In astrophysical applications of the EOS for hot stars and/or supernova
explosions it is commonly assumed that dense supernova matter can be
described well using a finite-temperature Hartree-Fock approximation applied
to an effective nucleon-\-nucleon interaction, and that the star consists
primarily of neutrons, protons, relativistic electrons and degenerate
electron neutrinos \cite{pines}.  However, for densities exceeding $2 - 3$
times nuclear saturation density, exotic phases of superdense matter are
possible; e.g., the interior of compact, astrophysical objects, such as
neutron stars, may be composed of strange quark matter (SQM)
\cite{bodmerwitten,glendenning,weber}.  Hitherto, however, such a scenario
remains speculative as there exists little evidence to support it or the
kindred hypothesis that domains of SQM might be formed under extreme
conditions of temperature and density~\cite{strangelets}.

Elucidating the composition of superdense, astrophysical objects requires a
knowledge of the EOS of strongly interacting matter at values of the
baryochemical potential close to that expected to induce a deconfinement
phase transition, and the plausibility of the results of any given study rest
on the accuracy of the EOS employed, which is difficult to judge {\it a
priori}.  Consequently, the exploration of alternative equations of state and
the identification of qualitatively consistent results is important.

Much existing research is based on the bag model EOS; e.g., \cite{weber}.
Other studies include those based on a string-flip, confining quark
interaction \cite{rbs}, which suggest that a phase of deconfined, massive
quarks is possible.  The absence of chiral symmetry is a defect common to all
these studies.  A covariant approach incorporating both confinement and
chiral symmetry is necessary and a step in that direction is presented in
\cite{drago}, which also argues that deconfinement can occur before chiral
symmetry restoration.  An EOS for quark matter at finite-$T$ is provided by
numerical simulations of lattice-QCD actions, however, finite chemical
potential continues to present difficulties~\cite{lks}.

The Dyson-Schwinger equations (DSEs) have been applied successfully to the
strong-interaction at $T=0=\mu$; i.e., to the study of confinement and
dynamical chiral symmetry breaking, and hadron
observables~\cite{anu98,tandy}.  The generalisation to finite-$(T,\mu)$ is
straightforward~\cite{prl,rapid,thermo,basti} and allows the simultaneous
study of the chiral and deconfinement phase transitions.  Herein we employ a
simple model~\cite{mn} whose bulk thermodynamic properties have recently been
elucidated~\cite{thermo}.  In many respects; e.g., the persistence of
nonperturbative effects into the quark-matter domain, the thermodynamic
properties of this model are qualitatively similar to those found in
numerical simulations of lattice-QCD.  In this note we present a first
exploration of the implications of this model for the stability and structure
of pure quark-matter stars.

This application requires only that we recapitulate a few important aspects
of the thermodynamic properties of our simple dynamical, confining model (DC
model)~\cite{thermo}.  In the confined domain the quark pressure is zero
because confinement does not allow any free quarks.  The $(T,\mu)$-dependent
bag pressure, which measures the difference between the pressure in the
Nambu-Goldstone phase and that in the Wigner phase:
\begin{equation}
{\cal B}(T,\mu) = P[S_{\rm NG}] - P[S_{\rm W}]\,,
\end{equation}
vanishes if the scalar piece of the quark self energy, $B(\tilde p_k)$,
becomes zero.  Chiral symmetry is manifest and the quark propagator has a
Lehmann representation when $B(\tilde p_k)=0$, which means we have chiral
symmetry restoration and deconfinement.  Hence the line ${\cal B}(T,\mu) = 0$
defines the phase boundary.

In the deconfined domain, each massless quark species $i=u,d$ contributes an
amount ($\tilde p_k = (\vec{p},\omega_k+i\mu)$, $\omega_k = (2 k +1)\pi T$,
$p=|\vec{p}|$)
\begin{eqnarray}
\label{pwigner}
P_{i}(T,\mu_i) &=& \frac{2 N_c}{\pi^2} T \sum_{k=0}^{\infty}
        \int_0^{\infty}~dp~p^2~
        \left\{\ln\left| \beta^2 \tilde p_k^2 \hat C^2(\tilde p_k)\right|
        - 1 +  {\sf Re}\left(\frac{1}{\hat C(\tilde p_k)}\right)  \right\},\\
\hat C(\tilde p_k)  & = &
\sfrac{1}{2}\left( 1 + \sqrt{1 + \sfrac{2 \eta^2}{\tilde p_k^2}}\right),
\end{eqnarray}
to the pressure, which is normalised to zero at the phase boundary.  Here
$\eta\approx 1\,$GeV is a mass-scale set~\cite{mn} by requiring a good
description of $\pi$ and $\rho$ masses.  In calculating this pressure a
modified free particle dispersion relation arises: ${e}(p;\mu_i)=
\kappa(p;\mu_i) + p$.  It is characterised by a new, dynamically-determined
energy-scale $\kappa(p;\mu_i)$, with $\kappa(0,0)\approx 0.6$ GeV, which
appears because even in the deconfined phase the vector piece of the
dressed-quark self energy is not trivial; i.e., ${\hat C}({\tilde p}_k)\neq
1$ in Eq. (\ref{pwigner}).  $\kappa(0,0)$ is a single, characteristic,
nonperturbative scale that measures the deviation of $\hat C({\tilde p}_k)$
from $1$ and its slow approach to the asymptotic limit: ${\hat C}({\tilde
p}_k)=1$.  The quark pressure in the Wigner phase is therefore
(${e}_\pm(p,\mu_i):= {e}(p,\mu_i)\pm \mu_i$)
\begin{eqnarray}\label{mnmodel}
P_{i}(T,\mu_i) = && \frac{N_c}{\pi^2}\,T\,\int_0^{\infty}dp~p^2~
\left\{\ln\bigg[1 + {\rm e}^{ -\beta \,{e}_-}\bigg] + \ln\bigg[1 + {\rm e}^{
-\beta\, {e}_+}\bigg]\right\}~.
\end{eqnarray}

The persistence of nonperturbative effects in the deconfined domain at
finite-$T$ agrees qualitatively with the findings of lattice-QCD
simulations~\cite{engels}.  It entails that the ultrarelativistic limit is
reached slowly.  Therefore, it is important to keep all three dressing
functions in the propagator and model their dependence on their arguments in
a qualitatively accurate manner.

For the following numerical calculations we use the EOS in
Eq.~(\ref{mnmodel}) with the simplification of neglecting the
$\mu$-dependence of $\kappa$, which reduces the computational effort
significantly.  As illustrated in \cite{thermo}, this is a good approximation
for $\mu< 3 \mu_c$.  (In our model, $\mu_c(T=0)\approx 0.28\,\eta$ and
$\mu_c(T) \approx\,$constant for $T<0.6\,\eta$.)  We will refer to this
equation of state as EOS$^{\rm DC}$.

The total pressure of nonstrange quark matter in this model was calculated in
\cite{thermo}.  It is obtained as the sum of the u- and d-quark contributions
\begin{equation}
\label{mnpress}
P^{\rm DC}(T,\mu_u,\mu_d)=P_u(T,\mu_u)+P_d(T,\mu_d)
\end{equation} 
and is depicted in Fig.~4 therein.  The inclusion of strange quarks awaits an
answer to the question of whether the simple {\it Ansatz} for the gluon
propagator used in \cite{thermo} can model $3$-flavour QCD accurately.  It is
not necessary for our present, exploratory purposes and its qualitative
impact can be anticipated.

Our model is characterised by deconfinement and chiral symmetry restoration
as a dynamical result following the introduction of the external mass-scales:
$(T,\mu)$, and, as noted above, the {\it quark pressure} as a thermodynamic
potential, is {\it zero} in the {\it confined domain}.  Consequently the
contribution of quarks to all other thermodynamic quantities: entropy
density, number densities, etc., as derivatives of this potential
with-respect-to the set of thermodynamic variables $(T, \mu)$, also vanishes.
In the confinement domain all the thermodynamic quantities are determined by
hadronic degrees of freedom.

As a means of elucidating the consequences of this model we compare our
results with the oft employed bag-model (BM), in which the EOS is
\begin{eqnarray}\label{bagmodel}
P^{\rm BM}(T,\mu_u,\mu_d) & = &
P_{u}^{\rm UR}(T,\mu_u)+P_{d}^{\rm UR}(T,\mu_d) - B_P\,.\\
\label{pur}
P_{i}^{\rm UR}(T,\mu_i)&=&\frac{g_i}{24\pi^2}\bigg(\mu^4_i+2\pi^2\mu^2_iT^2+
\frac{7}{15}\pi^4T^4\bigg)
\end{eqnarray}  
is the pressure of a massless, ultrarelativistic gas for a fermion of
species $i$, $i=u, d, e$.  For the bag-model we have $N_c= 3$ and $g_{i} =
2\,N_c\,,\;i=u,d$, and the bag constant, $B_P$, is $(T,\mu)$-independent.  It
is introduced by hand to play the role of an external pressure necessary to
``confine'' the (ideal) Fermi gas of quarks, and its $(T,\mu)$-independence
entails a strong first-order deconfinement transition at all values of
$(T,\mu)$.\footnote{This is at odds with the expectation that the
deconfinement transition is second order at
$\mu=0$~\protect\cite{prl,thermo}.  Also, since the bag surface breaks chiral
symmetry explicitly, the chiral symmetry restoring transition is outside the
scope of the model.}  Herein we use $B_P = 57$ MeV/fm$^3$ \cite{weber} and
refer to this equation of state as EOS$^{\rm BM}$.

Using the two different equations of state, Eqs.~(\ref{mnpress}) and
(\ref{bagmodel}), it is straightforward to obtain all the thermodynamic
quantities: the partial densities $n_i=\partial P/\partial \mu_i$, the
entropy density $s=\partial P/\partial T$, the energy density $\varepsilon =
-P + Ts + \sum \mu_i n_i$, etc.  The only modification of the pressure
necessary for the case of neutron star matter in $\beta$-equilibrium is, in
both models, to add the contribution of the electron component, which for
massless-$e^-$ is given by
\begin{equation}
P_{e}(T,\mu_{e})= P_{e}^{\rm UR}(T,\mu_{e})\,, \; g_{e}=2\,.
\end{equation}

\begin{figure}[t]
\centerline{\epsfig{figure=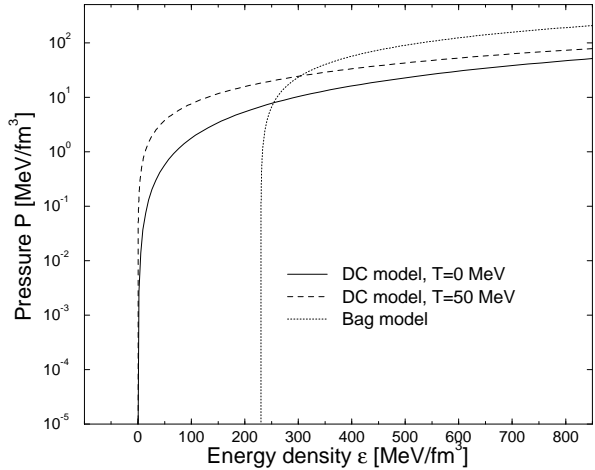,height=9.0cm,angle=-90}}
\caption{Total pressure of quark matter as a function of the energy density
for the DC-model at $T=0$ (solid line) and at $T=50$ MeV (dashed line). The
dotted line for the massless bag-model is valid at all
temperatures.\label{pe}}
\end{figure}
Figure~\ref{pe} depicts the quark pressure as a function of the energy
density in our model and in the bag model.  The essential differences are
clear.  In bag-model-like approaches the energy density has a large finite
value when the total pressure is zero as a direct consequence of the assumed
$(T,\mu)$-independence of $B_P$.  In our case, the energy density at the phase
boundary in the deconfined domain, defined by $\mu_c(T)$, is negligibly small
over the temperature interval $0\le T\le 50$ MeV, which is the domain
relevant to the problem of quark stars.  This feature is a direct
manifestation of the $(T,\mu)$-dependence of ${\cal B}$, which provides a
more realistic representation of the quark-dynamics of the phase transition.
One also observes the markedly slower approach to the ultraviolet limit in
our model.  This means that EOS$^{\rm BM}$ is much stiffer than EOS$^{\rm
DC}$, whose softness is consistent with expectations arising from numerical
simulations of lattice-QCD.

Most relativistic studies of pure quark-matter stars have been performed for
matter at $T\ll T_F$, where $T_F$ is the Fermi temperature.  However, in the
early stages of the life of a neutron/quark star; e.g., just after a
supernova explosion, an initial temperature $T\sim T_F $ is possible.  The
cooling of the star via neutrino emission requires minutes \cite{burrows},
which is a time-scale much greater than that required to establish
$\beta$-equilibrium.  It is therefore important to determine whether for
temperatures $T\sim T_F$ the stability criterion (Chandrasekhar limit) for
such proto-stars requires modification.

Other studies have assumed $T=\,$constant across a hot quark star
\cite{weber}, although an internal heating mechanism necessary to effect an
isothermal distribution is not elucidated.  This simplification leads to the
conclusion that the maximum mass of a quark star is increased by a few
percent when the temperature is increased from $T=0$ to $50\,$MeV.  However,
we argue that only local thermal and hydrodynamical equilibrium can reliably
be assumed in the early stages of the evolution of a quark star, and hence
that the temperature profile must vary over the star.

For clarity we enumerate the assumptions employed in our
calculations:\vspace*{-0.5\baselineskip}

(i)~The quark star is a spherically symmetric, compact object, in which the
matter is in local hydrodynamical and thermodynamical equilibrium with the
self-consistently determined gravitational field.  The pressure profile is
therefore determined by a solution of the Tolman-Oppenheimer-Volkoff equation
\begin{eqnarray}
\label{tov}
\frac{dP(r)}{dr}=-G(\varepsilon(r) + P(r))
\frac{m(r)+ 4\pi P(r) r^3 }{ r( r - 2 G m(r))}~,
\end{eqnarray}
where $G$ is the gravitational constant and $m(r)$ is the mass accumulated 
inside a distance $r$ from the center of the star, defined by
\begin{eqnarray}
\label{mr}
\frac{dm(r)}{dr}&=&4\pi r^2 \varepsilon(r) ~.
\end{eqnarray}

(ii)~ The central baryon number densities are higher than the densities
assumed to correspond to the confinement-deconfinement phase transition: $ n
> 2-3\,\, n_{0} $, where $n_0=0.17$ fm$^{-3}$ is the nuclear saturation
density, and the central temperatures $ T < 20 - 50$ MeV are typical of
neutron stars newly formed in a supernova explosion \cite{burrows,prakash}.
The radius of the star $R$ is obtained from the condition $ P(R)=0$ and the
total mass is then $ M=m(R)$.

(iii)~All the components of the hot matter (quark and lepton species:
$i=u,d,e^-$) are in chemical equilibrium with respect to the $\beta$-decay
process: $d + \nu_e \leftrightarrow u + e^-$, such that
\begin{equation}
\label{betaeq}
\mu_e - \mu_{\nu_e} =\mu_d-\mu_u > 0~.
\end{equation}
Hereafter we set $\mu_{\nu_e}=0$, which assumes the neutrinos escape quickly
from the star.

(iv)~The conditions of charge neutrality: $n_e= \left( 2 n_u - n_d \right)/3$
and the conservation of baryon number
\begin{equation}
\label{nr}
n = \sfrac{1}{3}\left(n_{u}+n_{d}\right)
\end{equation}
are fulfilled.  Thermodynamical consistency requires that the thermodynamic
functions obey Gibbs' law, which for conserved baryon number is
\begin{equation}
  \label{firstl}
  d\bigg(\frac{\varepsilon}{n}\bigg)-T d\bigg(\frac{s}{n}\bigg)+P 
  d\bigg(\frac{1}{n}\bigg)=0\,.
\end{equation}
The entropy per particle remains constant with respect to volume-deformations
under the forces of gravity such that the second term in Eq.~(\ref{firstl})
vanishes and the star profile is determined by 
\begin{equation}
  \label{adb}
  \frac{s}{n}\bigg(T(r),\mu(r)\bigg)= \frac{s}{n}\bigg(T(0),\mu(0)\bigg)~.
\end{equation}
\vspace*{-\baselineskip}

With these assumptions we have specified that complete set of equations:
(\ref{tov})-(\ref{adb}), necessary to determine the structure of a star with
given central temperature and chemical potentials once the EOS of the hot and
dense matter is specified.  We have employed them to obtain nonstrange quark
star configurations using EOS$^{\rm DC}$ and also EOS$^{\rm BM}$ for
comparison.

\begin{figure}[t]
\centerline{\epsfig{figure=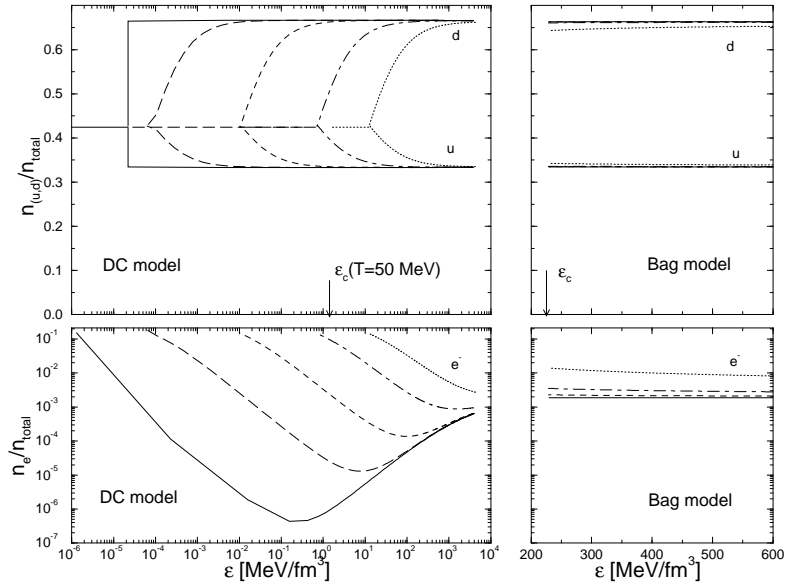,height=10.0cm,angle=-90}}
\caption{Composition of quark matter star in $\beta$-equilibrium with
electrons: DC-model (left panels); bag-model (right panels). Temperatures (in
MeV): $T=0$ (thin solid lines), $T=1$ (long dashed lines), $T=5$ (dashed
lines), $T=20$ (dot-dashed lines) and $T=50$ (dotted lines).  $e^-$ (lower
panels), and $u$- and $d$-quark (upper panels) concentrations are shown as a
function of the energy density.  The critical energy densities,
$\varepsilon_c$, for the bag-model are temperature independent.  In the
DC-model they are negligibly small and only $\varepsilon_c(T=50$ MeV) can be
shown in the figure. \label{composition}}
\end{figure}
An important consequence of the difference between these equations of state
highlighted in Fig.~\ref{pe}, is a strong modification at low densities of
the composition of quark matter in $\beta$-equilibrium with electrons,
depicted in Fig.~\ref{composition}.  Due to the low critical energy density
in our model; i.e, the energy density at deconfinement (e.g.,
$\varepsilon_c(T=50\,{\rm MeV}) = 1.4$ MeV/fm$^{3}$), we observe a transition
to isospin-symmetric matter as the energy density is decreased, whereas in
the bag model, with $\varepsilon_c = 4 B_P= 228 $ MeV/fm$^{3}$, the partial
fractions of the quark and lepton species are little changed over the
relevant energy domain.

\begin{figure}[t]
\centerline{\epsfig{figure=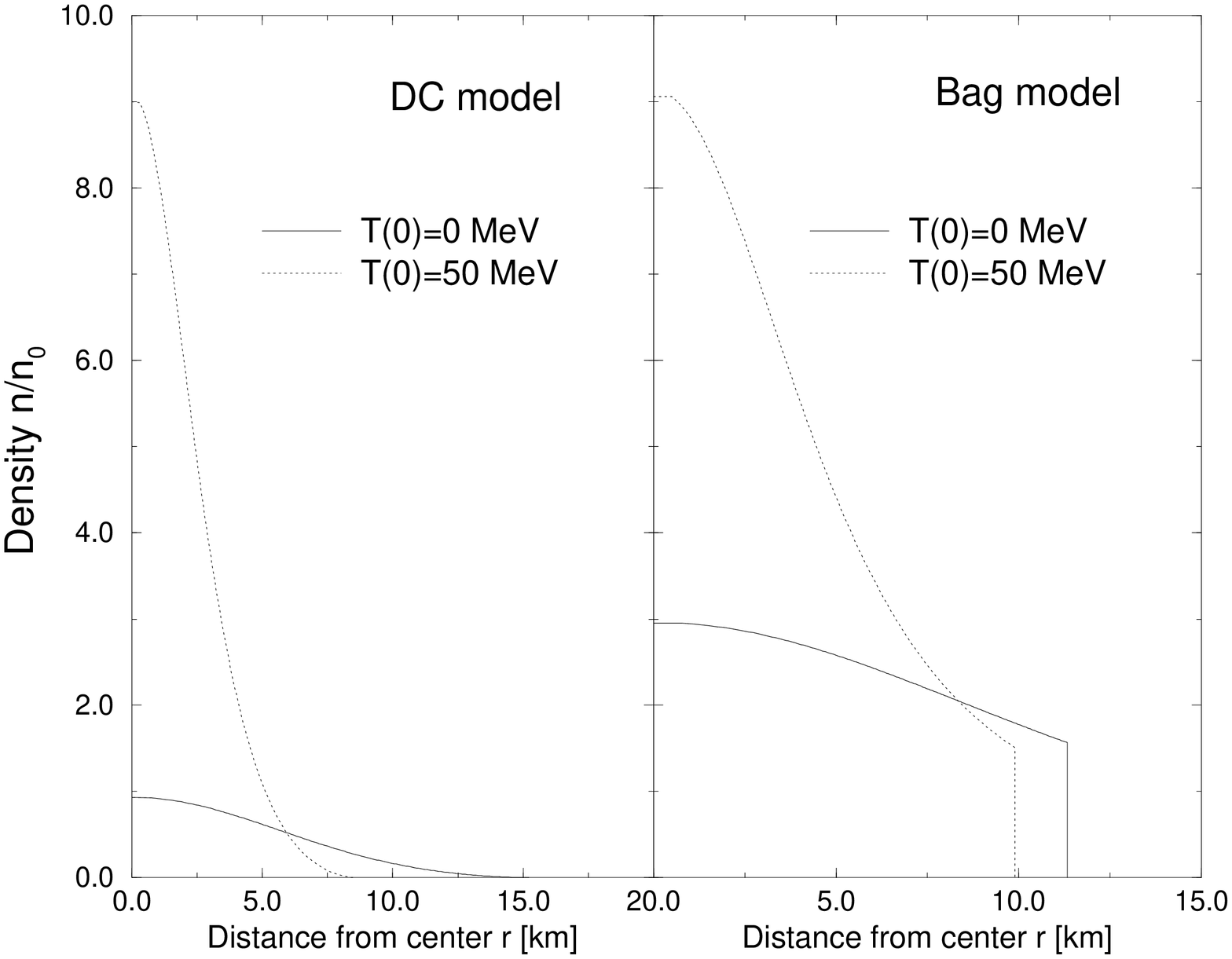,height=8.0cm,angle=-90}}
\centerline{\epsfig{figure=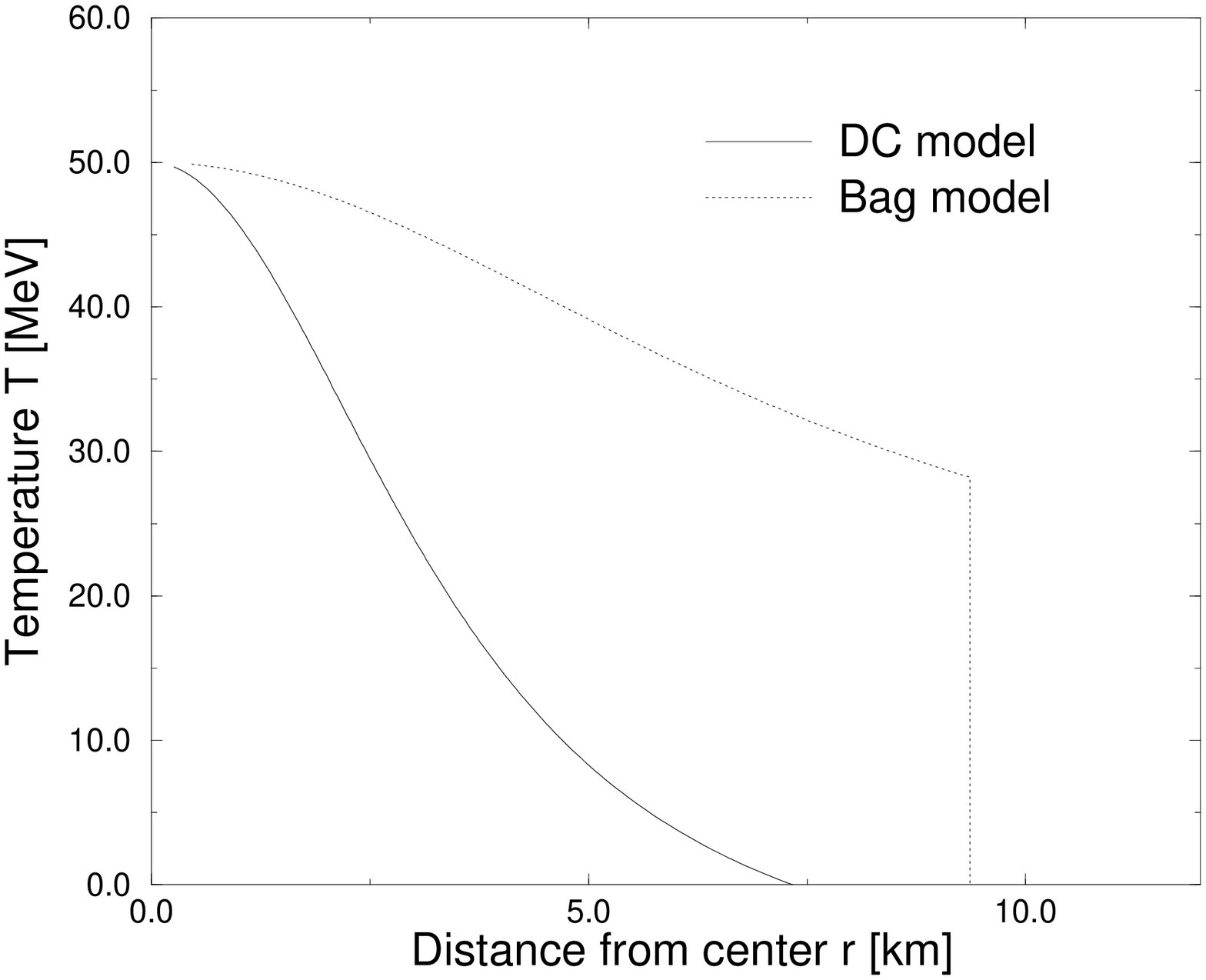,height=8.0cm,angle=-90}}
\vspace*{\baselineskip}
\caption{Upper Figure -- Density profiles for adiabatic star configurations:
DC-model (left panel); bag-model (right panel), for central temperatures
$T(0)=0$ (solid lines) and $T(0)=50$ MeV (dotted lines).  For each model the
total baryon number is kept fixed: $N^{\rm DC}= 0.866~ N_{\odot}$ and $N^{\rm
BM}= 5.766~ N_{\odot}$. Lower Figure -- Temperature profile for adiabatic
quark star configurations with $T(0)=50$ MeV and central density $n(0)= 7.2~
n_0$.\label{structure1}}
\end{figure}
In Fig.~\ref{structure1} we plot the density and temperature profiles for a
quark star with fixed baryon number:
\begin{equation}
N = 4\pi \int_0^R\,dr\,r^2\, 
\frac{n(r)}{[1 - 2 \,m(r)\,G/r]^{\sfrac{1}{2}}}\,,
\end{equation}
where $m(r)$ is the mass distribution, Eq.~(\ref{mr}), and $n(r)$ is the
baryon-number density, Eq.~(\ref{nr}).  Our reference measure for $N$ is
$N_\odot := M_\odot/m_N$, with $m_N$ the nucleon mass.  These figures provide
a further comparison between the two models.  EOS$^{\rm DC}$ leads to
profiles that approach zero smoothly at the surface of the star, whereas
using EOS$^{\rm BM}$ the density and temperature are nonzero at the surface.
This is a direct consequence of the difference between constant $B_P$ and
$(T,\mu)$-dependent ${\cal B}$, and how they characterise the phase
transition.  For a fixed central density, since the density profile vanishes
at the surface, the total number of particles in our model is less.  The
calculated difference in surface temperature means that the luminosity of a
star described by EOS$^{\rm DC}$ is less than that of a quark-matter star
described by EOS$^{\rm BM}$.

\begin{figure}[t]
\centerline{\epsfig{figure=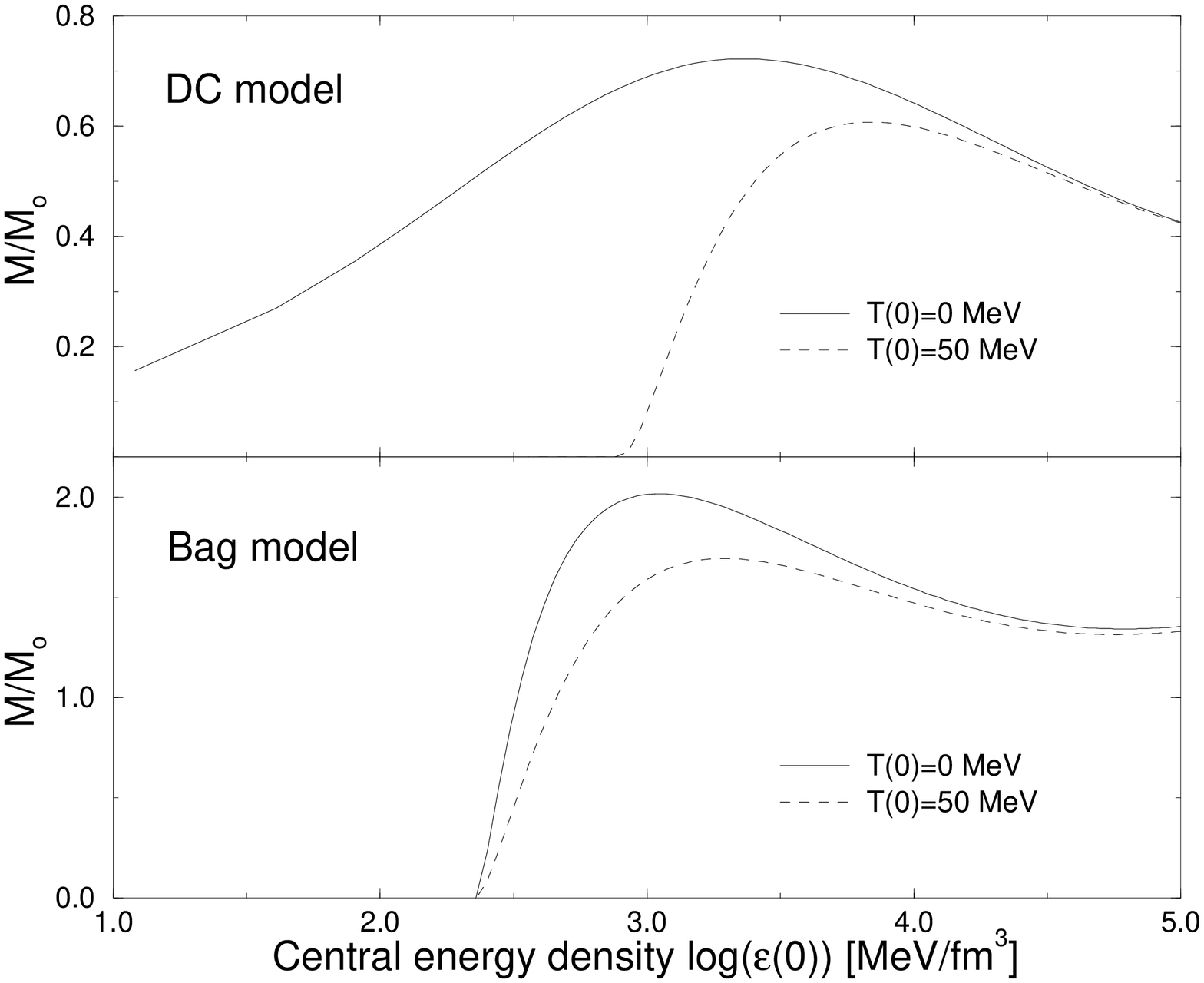,height=8.0cm,angle=-90}}
\centerline{\epsfig{figure=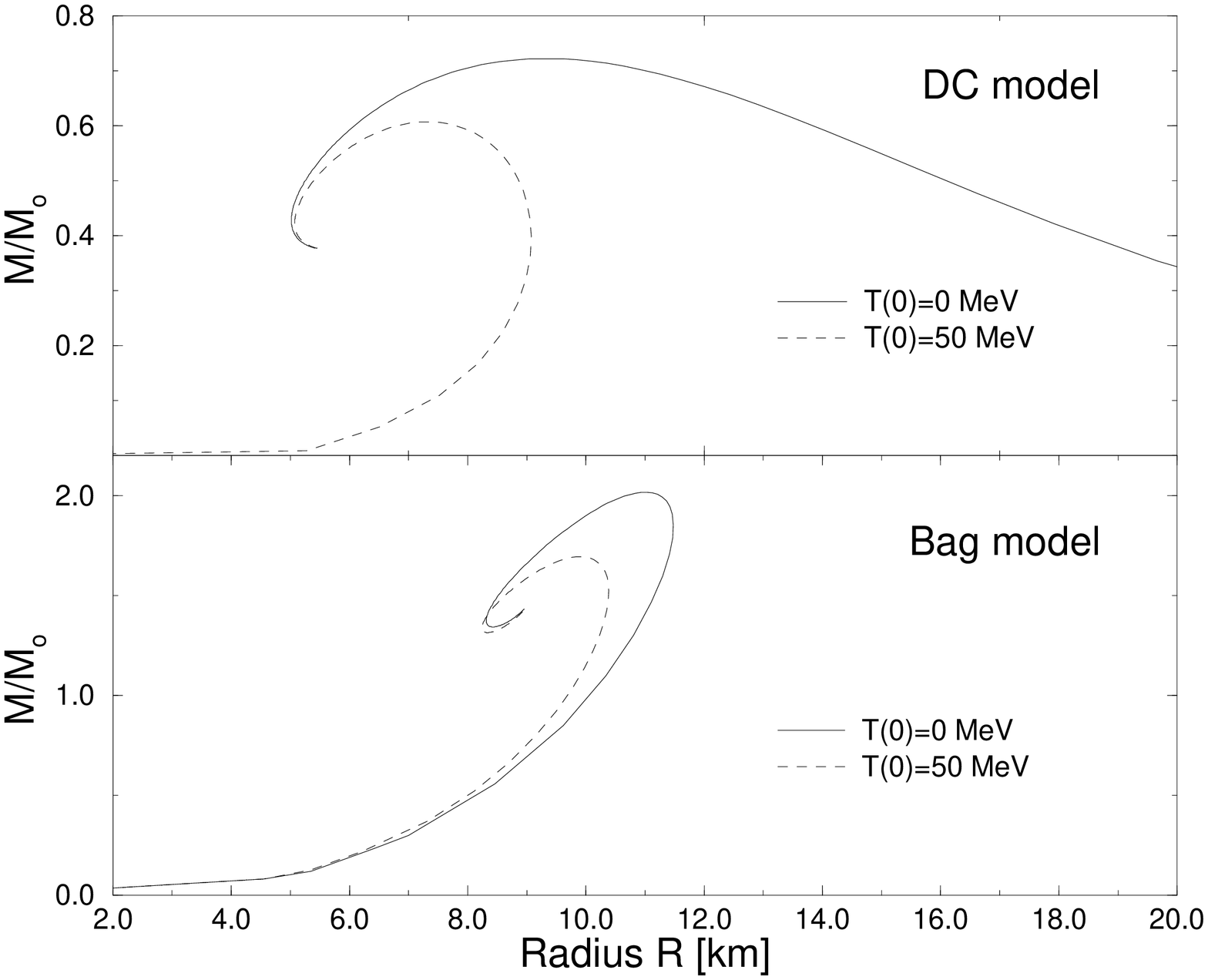,height=8.0cm,angle=-90}}
\vspace{\baselineskip}
\caption{Upper Figure -- Gravitational mass of quark stars as a function of
the central energy density: DC-model (upper panel); bag-model (lower panel).
At a central temperature of $T(0)=50\,$MeV (dashed lines) the maximum mass is
lowered by $\sim 20$\% when compared to the $T(0)=0$ case (solid lines).
Lower Figure -- Gravitational mass of quark stars as a function of the
radius.  At $T(0)=50$ MeV the objects are more compact than when $T(0)=0$.
\label{stability1}}
\end{figure}

Our analysis of the stability of quark stars is illustrated in
Fig.~\ref{stability1}.  At $T=0$ in the bag-model the maximum attainable mass
is approximately three-times more than that in our model.  This difference is
primarily the result of the fact that, when the pressure is zero, the energy
density in the bag-model is large whereas in our model it is small, as
depicted in Fig.~\ref{pe}.

In a real star the quark core is surrounded by a hadron shell, which will
contribute a finite amount to the total pressure.  In the bag-model the
$(T,\mu)$-independent bag constant {\bf mimics} this effect to some {\bf
indeterminable} extent.

In contrast, EOS$^{\rm DC}$ is obtained systematically and excludes by
construction the contribution of hadrons; i.e., it only describes the
contribution of quarks to the pressure.  Hadrons provide an additional,
additive contribution that is calculable in our framework; e.g, as proposed
in \cite{reg}.  The absence of this contribution in our present calculation
is the origin of the low maximum-possible quark-star mass in our model:
EOS$^{\rm DC}$ yields the maximum mass of a pure quark-matter star.

At finite-$T$ the maximum radius of a quark star is approximately the same in
both models: $R\approx 8\,$km in our model and $R\approx 10\,$km in the
bag-model.  The lower panel in Fig.~\ref{stability1} shows that increasing
$T$ leads to a reduction in both the maximum radius, $R$, and mass of a quark
star.  This is because increasing $T$ increases the compressibility; i.e., at
finite-$T$ the pressure increases less rapidly with density, and hence a
given gravitational mass occupies less volume.  This same effect, which
involves an increase in the central density, entails that the maximum mass of
a stable quark star is reduced, as depicted in Fig.~\ref{stability1}.

The star configurations shown in the upper panels of Fig. \ref{structure1}
can be considered as the initial and final states in the thermal evolution of
a star.  For a stable quark-star characterised by the quoted fixed total
baryon numbers we find the following masses
\begin{equation}
\begin{array}{lcc}
 T \,({\rm MeV})        &       50     &        0\\
{\rm DC:}&       0.55\, M_\odot &       0.54 M_\odot\,, \\
{\rm BM:}&       1.69 \, M_\odot &       1.62 M_\odot\,.
\end{array}
\end{equation}
Hence, in cooling, 2-4\% of the gravitational mass of a quark star is
radiated.

Our study demonstrates that the properties of a quark-matter star are
sensitive to the EOS and the temperature profile across the star, and to
emphasise this we recapitulate our main results.  Some are common to both
models and should therefore be reliable predictions: (i) When the temperature
$T(r)$ varies across the star the maximum allowable mass of a quark star is
reduced as the central temperature, $T(0)$, is increased.  Comparing $T(0)=0$
and $50\,$MeV, the reduction is $\sim 20$\%.  In contrast, if the temperature
is assumed to be constant across the star then the maximum allowable mass
increases slightly with increasing $T(0)$. (ii) At $T(0)=50\,$MeV the maximum
attainable radius of a pure quark star is $R \sim 8 - 10\,$km, with EOS$^{\rm
DC}$ giving the least upper bound.  (iii) In cooling a quark star radiates
2-4\% of its gravitational mass.

There are, however, significant disagreements and the predictions of both
models should be viewed cautiously in these cases: (i) Using EOS$^{\rm DC}$
the maximum mass of a pure quark-matter star is $M\sim 0.6-0.7\,M_{\odot}$,
which is approximately the same as the maximum mass of a star composed of an
ideal neutron gas.  It is only one-third of the maximum mass attainable using
EOS$^{\rm BM}$.  (ii) The temperature vanishes at the surface of an EOS$^{\rm
DC}$ star whereas in an EOS$^{\rm BM}$ star the surface temperature is
approximately 60\% of the central temperature. (iii) The energy density at
the surface of a quark star described by EOS$^{\rm DC}$ is less than that of
a star described by EOS$^{\rm BM}$.  Hence, in an EOS$^{\rm DC}$ star the
electron density fraction is higher and the quark matter is approximately
isospin-symmetric at the surface, whereas in an EOS$^{\rm BM}$ star the
composition remains unchanged for all relevant energy densities.

Our study is based on a particularly simple gluon propagator.  Those that use
a more sophisticated model~\cite{prl,basti} lead to qualitatively the same
behaviour of the quark propagator and hence to a similar EOS for pure
quark-matter.  We anticipate only small quantitative modifications, such as a
slight stiffening of the EOS and concomitant improvement of its
correspondence with lattice-QCD simulations.  Therefore our analysis provides
a qualitatively instructive first exploration of the properties of quark
matter when the EOS is calculated in a dynamical model that describes chiral
symmetry restoration and deconfinement simultaneously.

An important next step is the inclusion of hadronic bound states and their
contribution to the pressure, which is a systematic but complex extension.
Insofar as we have explicitly neglected hadrons, our equation of state is
only directly relevant to the idealised case of a pure quark-matter star, and
what might be considered as the physically unreasonable aspects of our
results are primarily the consequence of their omission.  This extension is
necessary if we are to make statements about the evolution of real stars.
Hence, at present, the mimicking of these effects in EOS$^{\rm BM}$, although
uncontrolled, makes it a firmer foundation for a phenomenology of real stars.
Nevertheless, incipient in EOS$^{\rm DC}$ is a constructive alternative that
does not contain hidden degrees of freedom and can be improved
systematically: EOS$^{\rm DC}$ holds the promise of an unaffected,
phenomenologically acceptable description of hot and dense nuclear matter.

%%%%%%%%%%%%%%%%%%%%%%%%%%%%%%
DB and SS acknowledge the hospitality of the Faculty of Theoretical Physics
at the Yerevan State University where this research began; HG, GP and CDR
that of the Physics Department at the University of Rostock where it was
completed; and we all acknowledge JINR-Dubna for its hospitality during the
workshop on {\em Deconfinement at Finite Temperature and Density} in October
1997.  This work was supported in part by Deutscher Akademischer
Austauschdienst; the US Department of Energy, Nuclear Physics Division, under
contract number W-31-109-ENG-38; the Hochschulsonderprogramm-III under
project no.~0037-6003; the National Science Foundation under grant
no.~INT-9603385; the Volkswagen Stiftung, grant no.~I/71 226; and benefited
from the resources of the National Energy Research Scientific Computing
Center.
%

%..............................Actual Figures

\end{document}